%% file: main.tex
\newif\ifarxiv
\arxivtrue  

\ifarxiv
    \documentclass[sigconf]{acmart}
\else
    \documentclass[sigconf,anonymous,review,natbib=true]{acmart}
\fi

\usepackage{enumitem}
\usepackage{multirow}

\ifarxiv
    \setcopyright{none}
    \renewcommand\footnotetextcopyrightpermission[1]{}
\else
    \copyrightyear{2027}
    \acmYear{2027}
    \acmConference[WSDM '27]
      {The 20th ACM International Conference on Web Search and Data Mining}
      {February 15--19, 2027}
      {Hong Kong SAR, China}
\fi

\begin{document}

\ifarxiv
    \title[Downstream Utility of Evidence-Aware Retrieval]
      {Assessing the Downstream Utility of Evidence-Aware Retrieval in RAG}
\else
    \title[Validity Does Not Compose]
      {Validity Does Not Compose: Does Evidence-Aware Retrieval Actually Improve RAG?}
\fi

\ifarxiv
    \author{Utshab Kumar Ghosh}
    \affiliation{%
      \department{Department of Computer Science}
      \institution{Missouri University of Science and Technology}
      \city{Rolla}
      \state{MO}
      \country{USA}
    }
    \orcid{0000-0003-3096-6909}
    \email{u.ghosh@mst.edu}

    \author{Debayan Mukhopadhyay}
\affiliation{%
    \institution{University of Calcutta}
    \city{Kolkata}
  \state{West Bengal}
    \country{India}
}
\email{debayan.mukherjee14@gmail.com}

    \author{Shubham Chatterjee}
    \affiliation{%
      \department{Department of Computer Science}
      \institution{Missouri University of Science and Technology}
      \city{Rolla}
      \state{MO}
      \country{USA}
    }
    \orcid{0000-0002-6729-1346}
    \email{shubham.chatterjee@mst.edu}

    \renewcommand{\shortauthors}{Ghosh, Mukhopadhyay, and Chatterjee}
\fi

\ifarxiv
    \input{abstract_arxiv}
\else
    \input{abstract}
\fi


\maketitle

\input{01_introduction}

\input{02_related_work}

\input{03_grounding_framework}

\input{04_benchmark_validity}

\input{05_optimization_validity}

\input{06_downstream_validity}

\input{07_intervention_validity}

\input{08_discussion}

\input{09_conclusion}

\input{10_ethics}

\balance
\bibliographystyle{ACM-Reference-Format}
\bibliography{references}
\end{document}

%% file: abstract_arxiv.tex
\begin{abstract}

Retrieval evaluation for retrieval-augmented generation (RAG) is increasingly
designed around whether retrieved passages contain evidence that can support
generation, rather than topical relevance alone. We study whether this closer
alignment with downstream evidence needs also makes retrieval evaluation more
useful for the decisions built from it.

Across five retrieval benchmarks and an end-to-end TREC RAG 2025 setting, we
examine an answer-support signal in four roles: comparing retrievers, guiding
retrieval training and system selection, predicting downstream answer quality,
and filtering the evidence supplied to a generator. The signal changes
retrieval rankings, but its downstream value is not uniform. It does not
reliably improve retriever training; the benefit of using it for system
selection depends on how the generator is instructed to use the retrieved
evidence; and retrieval scores based on it do not robustly predict answer
quality on unseen topics. In a direct evidence intervention, human annotators
confirm that filtering preferentially preserves passages containing useful
answer evidence, yet different answer evaluators reach different conclusions
about whether the resulting answers improve.

These results show that making retrieval evaluation more closely reflect the
evidence needed for generation does not by itself make every downstream use of
that evaluation more reliable. RAG evaluation methods should therefore be
assessed with respect to the particular comparisons, decisions, and conclusions
they are intended to support.

\end{abstract}

%% file: abstract.tex
\begin{abstract}

Retrieval-augmented generation (RAG) increasingly evaluates the retrieval component by
whether the retrieved passages contain evidence useful for answering the query, rather than
by topical relevance alone. This is a natural improvement: retrieval should
be judged by what downstream generation actually needs. But a more meaningful
evaluation signal does not necessarily make every decision based on that
signal more reliable.

We study this problem by following answer-support-aware evaluation through the
RAG pipeline. We ask whether it changes how retrievers are compared, provides
useful supervision for improving or selecting them, predicts downstream answer
quality, and supports reliable conclusions about whether better retrieved
evidence improves the final answer. Across five retrieval benchmarks and an
end-to-end TREC RAG 2025 setting, we find a consistent gap between making an
evaluation signal more closely reflect answer-bearing evidence and making that
signal useful for downstream decisions. Answer support changes retrieval
conclusions, but does not reliably improve training, its value for system
selection depends on the generation regime, and its scores do not robustly
predict answer quality on unseen topics. Even when human judgments confirm
that an intervention improves the retrieved evidence, the measured downstream
benefit depends on the answer evaluator.

We call this the \emph{validity-composition problem}: evidence that an
evaluation is meaningful for one use does not automatically validate the next
decision or conclusion built from it.

\end{abstract}

%% file: 01_introduction.tex
\section{Introduction}
\label{sec:introduction}

Retrieval-augmented generation (RAG) depends on retrieving evidence that can
support the generated answer, motivating evaluation criteria that go beyond
topical relevance. TREC RAG 2025~\cite{upadhyay2025trecrag}, for example,
distinguishes merely related passages from those that cover requested
sub-narratives, while RAGTIME 2025~\cite{lawrie2025ragtime} rewards information
useful for constructing the final report. This shift is also empirically
motivated: downstream-aware retrieval evaluation can align more closely with
RAG performance~\cite{salemi2024erag}, and utility-aware retrieval or evidence
selection can improve generation itself~\cite{qu2025upliftrag}. Together, these
developments suggest that an evaluation signal better aligned with
answer-bearing evidence may also provide a better basis for improving and
choosing retrievers.

That inference, however, has not been established. An evaluation criterion can
change which retrievers appear better without making the resulting comparison
a better basis for training, system selection, or predicting downstream
quality. Thus, showing that answer support is a more meaningful retrieval
criterion is distinct from showing that the decisions subsequently built from
it are empirically justified.

This motivates our central question:
\emph{If retrieval evaluation is better aligned with the evidence that RAG
needs, how far does that improvement carry through to the downstream decisions
built from it?}
We follow the same answer-support signal through a sequence of increasingly
consequential uses: comparing retrievers, using it for training and system
selection, predicting answer quality, and finally determining whether an
intervention that improves the retrieved evidence enhances the RAG system.

We find that the benefit does not propagate automatically. Answer-support-aware
evaluation changes retrieval conclusions, but using the same signal for
training does not reliably improve retrieval. The retrievers it favors improve
held-out answers under one generation regime but not another, and its retrieval
scores do not robustly predict answer quality on unseen topics. These failures cannot be dismissed as consequences of an arbitrary evidence signal: in our experiments, human annotators confirm that answer-support filtering gives the generator more answer-bearing evidence. We then evaluate the same generated answers with two different answer evaluators. Qwen finds that the filtering improves answer quality, whereas Claude finds essentially no improvement. Thus, even after verifying that the generator received better evidence, whether the resulting answers are judged to be better depends on how answer quality is measured.

Together, these findings expose what we call the
\emph{validity-composition problem}: evidence supporting an evaluation signal
for one use does not automatically support the next inference or decision
built from it. This is consistent with the broader measurement principle that
validity concerns particular interpretations and uses of a measure
~\cite{messick1995validity,kane2013validating}, but we show empirically how
this problem arises across the stages of a modern RAG pipeline.
The methodological implication is that the validity of an evaluation signal must be established with respect to the decisions it is used to support, not merely with respect to the construct it is intended to measure.

\smallskip 
\noindent 
\textbf{Contributions.} We make the following contributions: 
\begin{itemize}[leftmargin=*] 
    \item \textbf{We show when evidence-aware retrieval evaluation changes system conclusions.} Large disagreement with answer-bearing evidence is not sufficient: the effect becomes consequential when that mismatch is system-selective, affecting competing retrievers differently. 
    \item \textbf{We uncover a disconnect between better-aligned retrieval evaluation and downstream RAG decisions.} Answer-support-aware evaluation can materially change which retriever appears best without reliably improving optimization, held-out system selection, or prediction of answer quality. 
    \item \textbf{We show that this disconnect persists even after validating the evidence improvement with humans.} Independent annotators support the evidence distinction and confirm that answer-support filtering preferentially preserves evidence-bearing passages. Yet, with the generated outputs held fixed, different answer evaluators can lead to different conclusions about whether the same intervention improved answer quality. 
    \item \textbf{We identify the resulting validity-composition problem for RAG evaluation.} Our findings show that evaluation methods must be validated for the decisions they are intended to support, not only for the constructs they measure. This applies both to using retrieval metrics for system choice and optimization and to using answer evaluators to determine whether a RAG intervention worked. 
\end{itemize}

\smallskip
\noindent
Below, we follow this chain from evidence judgments to end-to-end conclusions. We test in turn whether answer-support-aware evaluation changes retrieval comparisons, whether those comparisons support better system decisions, whether retrieval scores predict answer quality, and whether a human-supported improvement in retrieved evidence yields a robust downstream conclusion.

%% file: 02_related_work.tex
\section{Related Work}
\label{sec:related_work}

\S\ref{sec:introduction} raises a tension: evidence-aware retrieval is a natural response to what RAG needs, yet a more meaningful signal need not support better downstream decisions. Prior work makes both sides plausible: evidence- and downstream-aware evaluation can improve RAG, while IR and measurement research show that a measure's usefulness depends on the inference or decision it supports.

\textbf{From Relevance to Evidence Utility in RAG.}
The move beyond topical relevance predates modern RAG evaluation. EXAM evaluates retrieve-and-generate systems through questions that the returned information should enable users to answer~\cite{sander2021exam}, while subsequent work develops question-, nugget-, and rubric-based measures of useful information~\cite{dietz2024workbench,farzi2024pencils}. RAGAS and ARES similarly separate properties of retrieved context from faithfulness and answer quality~\cite{es2024ragas,saadfalcon2024ares}. More recently, TREC RAG 2025 rewards evidence that contributes to requested sub-narratives, and RAGTIME 2025 distinguishes answer-useful information from material that is merely topical~\cite{upadhyay2025trecrag,lawrie2025ragtime}.

This progression reflects a broader IR insight: individually relevant results need not form a useful result set. Diversity-aware evaluation, for example, rewards coverage of distinct aspects rather than repeated evidence about the same aspect~\cite{clarke2008novelty}. RAG sharpens this distinction because retrieved material is not the final product; it is evidence supplied to a generator. eRAG makes this downstream role explicit by evaluating retrieved documents through their effect on generation and reports stronger alignment with RAG performance than conventional relevance judgments~\cite{salemi2024erag}. Such results make the expectation studied in this paper plausible: evaluation designed around answer-useful evidence can be more informative than relevance alone.

\textbf{When Better Retrieval Leads to Better Generation.}
Other work shows that downstream-aware retrieval signals can sometimes be acted upon successfully. Uplift-RAG defines document utility through marginal contribution to generation and uses that signal to improve reranking and evidence selection~\cite{qu2025upliftrag}. Lee et al.~\cite{lee2025inferencescaling} similarly estimate passage utility with respect to generator behavior, while sub-question coverage has been used to measure and improve whether retrieved evidence addresses important parts of an information need~\cite{xie2025subquestion}.

At the same time, better retrieval does not determine better generation by itself. RGB shows that generators differ in their ability to use noisy, incomplete, or conflicting evidence~\cite{chen2024rgb}; RAGGED and related work report interactions among retriever, reader, and context size~\cite{hsia2025ragged,vladika2025context}; and recent work distinguishes retrieval utility from final answer quality rather than treating them as interchangeable~\cite{tian2026predicting}. Thus, downstream-aware retrieval signals \emph{can} improve RAG, but their value depends on how the retrieved evidence is used. What remains unclear is whether a signal that better captures answer-useful evidence also becomes a better objective for optimization, a better basis for system selection, or a useful predictor of answer quality.

\textbf{From Evaluation Reliability to Decision Validity.}
IR evaluation has long separated properties of judgments from properties of the system conclusions drawn from them. Assessor disagreement, incomplete pools, topic sampling, and metric choice can all affect effectiveness estimates~\cite{zobel1998reliable,voorhees2000variations,buckley2000stability,buckley2004incomplete,sanderson2005effort}. Importantly, substantial disagreement among assessors can coexist with stable relative system rankings~\cite{voorhees2000variations}. The converse is also important for our setting: changing judgments or a leaderboard does not establish that the new ordering is a better basis for choosing systems.

The growing use of LLMs for relevance assessment makes this distinction particularly salient. Faggioli et al.\ caution against treating agreement with existing human assessments as sufficient justification for replacing them with LLM judgments~\cite{faggioli2023perspectives,faggioli2024who}. Clarke and Dietz similarly argue that reproducing human relevance judgments does not establish that LLM-generated qrels are safe reusable evaluation targets~\cite{clarke2025replace}, while Dietz et al.\ articulate broader principles for determining when LLM judges are appropriate~\cite{dietz2025principles}.

Measurement theory provides the general principle behind these concerns: validity supports a particular interpretation or use of a measure, not the measure in isolation~\cite{messick1995validity,kane2013validating}. In Kane's argument-based framework, each additional inference or decision requires supporting evidence. Applied to RAG, showing that answer support captures a meaningful evidence distinction does not establish that it is also a useful optimization objective, selection criterion, or predictor of downstream quality. These uses must be tested separately.

\textbf{Evaluator Dependence in RAG Conclusions.}
The same issue arises at the other end of the pipeline, where generated answers become the objects of evaluation. LLM judges exhibit systematic biases and sensitivity to evaluation conditions~\cite{zheng2023mtbench}. Recent work therefore examines consequences beyond item-level agreement. JuStRank studies LLM judges as system rankers and shows that aggregation can expose biases not apparent from individual judgments~\cite{gera2025justrank}; the Progress Illusion shows that strong aggregate meta-evaluation can conceal poor discrimination among systems of similar quality~\cite{xu2025progress}. RAGTIME reports differences between automatic and human evaluation of RAG runs~\cite{lawrie2025ragtime}, while Auto-Judge studies reliability and vulnerabilities of judges for citation-grounded RAG~\cite{farzi2026autojudge}.

For system development, however, the consequential question is often not whether two evaluators assign the same scores, but whether they support the same decision. If the generated outputs are fixed, score disagreement is unsurprising. More consequential is disagreement about whether an intervention improved the system. Evaluator robustness should therefore be examined not only through agreement on individual outputs or system rankings, but also through the stability of the substantive conclusion drawn from an experiment.

\smallskip
\noindent
This section establishes that both evidence- and downstream-aware signals can produce genuine improvements, while evaluation theory cautions that success for one use does not establish success for another. What remains unresolved is how far an improvement in evidence evaluation actually carries: \emph{when retrieval evaluation becomes better aligned with answer-bearing evidence, which downstream decisions does that improvement support?}

%% file: 03_grounding_framework.tex
\section{From Evidence Judgments to RAG Decisions}
\label{sec:framework}

We use \emph{answer support} as a concrete probe of evidence-aware retrieval
evaluation. A passage is \emph{answer-supporting} when it contains evidence
that contributes directly to answering the information need, rather than
merely discussing the same topic. To assess this distinction, we first
decompose each passage into atomic, self-contained claims using Gemma3-27B.
An answer-support judge (GPT 4.1) evaluates these claims against the information need
and assigns a single 0--3 grade to the query--passage pair. Grade 0 indicates
no answer-supporting evidence, 1 means partial or insufficient
evidence, and 2--3 satisfy the answer-support criterion. 

\begin{figure}[t]
    \centering
    \includegraphics[width=\linewidth]{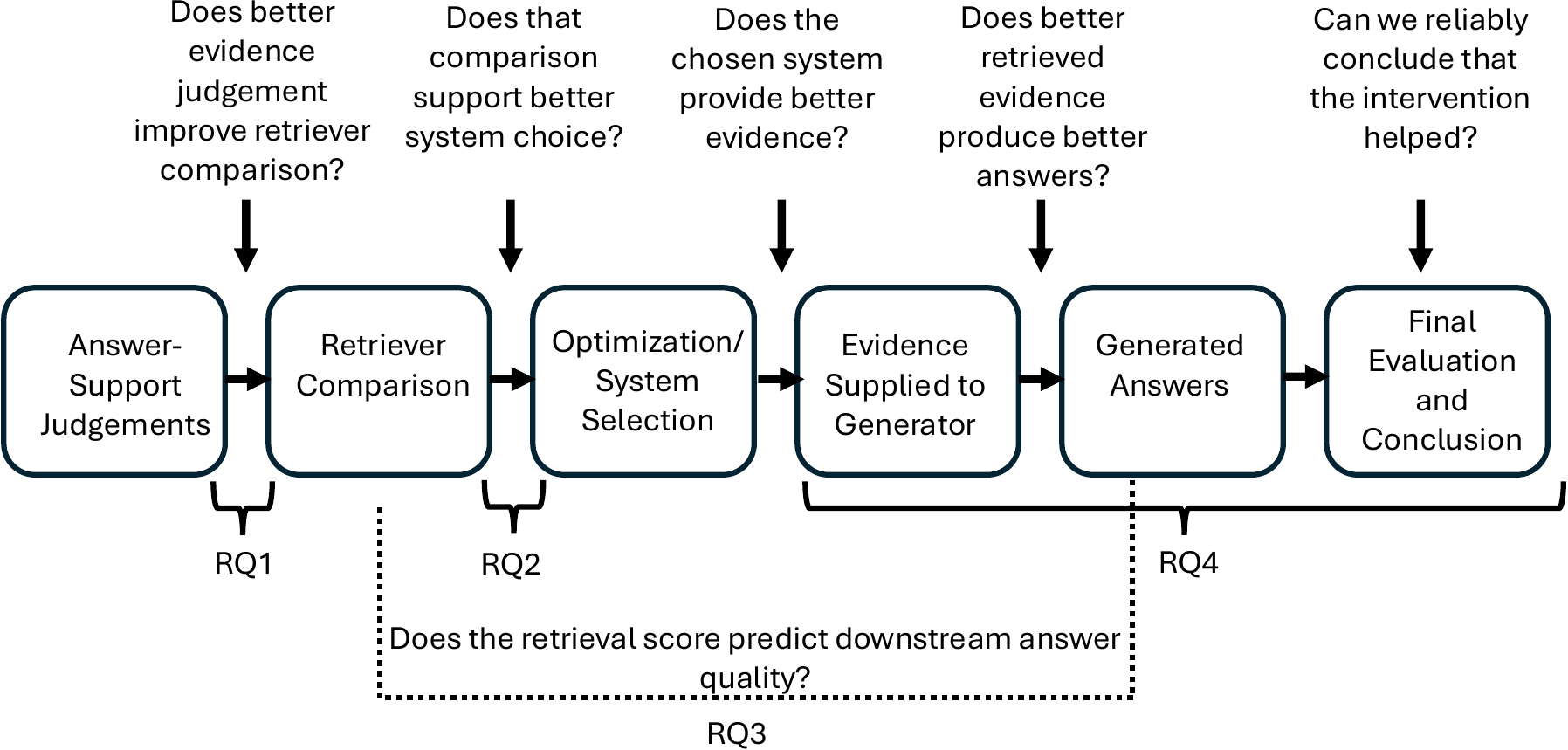}
    \caption{From answer-support judgments to end-to-end RAG conclusions.
    Each arrow represents a distinct empirical claim tested by RQ1--RQ4.
    The dashed RQ3 link represents the cross-stage claim that retrieval
    evaluation should predict downstream answer quality.}
    \label{fig:validity_chain}
\end{figure}

The central question of this paper is what follows once such judgments are
available. They can be used to change how retrievers are evaluated, to guide
system selection, to construct retrieval scores
treated as proxies for downstream answer quality, or to determine which
retrieved evidence reaches the generator. These are different uses of the
same underlying signal, and each supports a different claim about what that
signal can tell us. Figure~\ref{fig:validity_chain} follows these claims from
retrieval evaluation to the final RAG conclusion.

A meaningful answer-support judgment does not guarantee that these downstream
uses are also justified. Changing the evidence criterion may change which
retriever appears better without making that comparison a better basis for
optimization or system selection. A retrieval score may align more closely
with downstream performance without predicting answer quality on unseen
topics. And even an intervention that supplies better evidence does not by
itself establish that the resulting answers improved if that conclusion
depends on how answer quality is evaluated.

We call this the \emph{validity-composition problem}: evidence supporting an
evaluation signal for one use does not automatically validate the next
inference or decision built from it. This is RAG-specific shorthand for the
established principle that validity concerns particular interpretations and
uses of a measure~\cite{kane2013validating}. The relevant question is
therefore not whether an evaluation signal is ``valid'' in isolation, but
which uses of that signal the available evidence actually supports.

\smallskip
\noindent
\textbf{Research Questions.}
We test four links in this chain:

\textbf{RQ1: When does answer-support-aware evaluation change retrieval
conclusions?}
We test when answer support changes retriever rankings, whether those changes
exceed comparable random perturbations of the judgments, and why some
benchmarks are more affected than others.

\textbf{RQ2: Does answer support provide a better basis for optimizing or
selecting retrievers?}
We test whether using answer support as retrieval supervision improves
retrieval and whether retrievers selected by answer-support-aware evaluation
produce better answers on held-out topics.

\textbf{RQ3: Do answer-support-aware retrieval scores predict answer quality?}
We test whether these scores provide reliable out-of-sample information about
downstream RAG performance.

\textbf{RQ4: If the retrieved evidence improves, can we reliably conclude that
the RAG system improved?}
We use answer-support judgments to change the evidence supplied to the
generator, verify with human judgments that the resulting context contains
more answer-bearing evidence, and test whether the conclusion that the answers
improved is robust to evaluator choice.

%% file: 04_benchmark_validity.tex
\section{When Evidence-Aware Evaluation Matters}
\label{sec:retrieval_conclusions}

We begin with the first link in Figure~\ref{fig:validity_chain}: when does
answer-support-aware evaluation change how retrievers are compared? We test
this by evaluating the same retrieval runs under conventional relevance and
under a criterion that rewards answer-supporting evidence.

\smallskip
\noindent
\textbf{Experimental setup.}
We use five test sets. These include three BEIR~\cite{thakur2021beir} collections: TREC-COVID~\cite{voorhees2020treccovid}
(50 queries), NFCorpus~\cite{boteva2016} (323), and
SciFact~\cite{wadden-etal-2020-fact} (300), together with
TREC-DL 2019~\cite{craswell2020overviewtrec2019deep} (43) and
TREC-DL 2020~\cite{craswell2021overviewtrec2020deep} (54).
For the three BEIR collections, we evaluate 10 systems: BM25,
SPLADE-v3~\cite{lassance2024spladev3newbaselinessplade},
BGE~\cite{chen2025m3embedding}, and
Contriever~\cite{izacard2022unsupervised} as first-stage retrievers,
together with six rerankers over the BM25 top-100. The rerankers are three
cross-encoders (MiniLM-L6, MiniLM-L12, and BGE-reranker-base) and three
bi-encoders (BGE, E5~\cite{wang2024textembedding}, and
GTE~\cite{li2023generaltext}). For TREC-DL 2019 and 2020, we evaluate
11 systems: BM25, SPLADE-v3, BGE, TCT-ColBERT~\cite{lin2020distilling},
and TAS-B~\cite{hofstatter2021tasb}, together with the same six rerankers.
First-stage runs use Pyserini.
NFCorpus has one retrieval-coverage asymmetry. BM25 returns no results for
15 of its 323 queries, so BM25 and the six BM25-seeded rerankers cover
308 queries, whereas the three neural first-stage systems cover all 323.
We retain the full 323-query evaluation set, treating a system that returns
no results for a query as receiving zero effectiveness on that query.

\textbf{\textit{Constructing answer-support-aware qrels.}}
We use answer support as a stricter criterion on
passages already judged relevant, rather than constructing a new judgment pool
from scratch. We first define the original positive set using each benchmark's
native relevance threshold: grade $\geq 1$ for BEIR and grade $\geq 2$ for
TREC-DL, and then apply the answer-support procedure from
\S\ref{sec:framework} to those positives. This design isolates the effect of
requiring originally relevant passages to also contain answer-bearing
evidence: an original positive either retains its benchmark gain if its
answer-support grade is at least 2 or receives zero gain otherwise. We do not
add newly discovered positives from previously nonrelevant or unjudged
passages. This also gives a clean matched-random counterfactual in which the
same number of original positive judgments can be removed at random, allowing
us to ask whether the particular positives rejected by the answer-support
criterion alter system conclusions more than comparable judgment attrition.


\textbf{\textit{Comparing retrieval conclusions.}}
We then hold every retrieval run fixed and score the same rankings under the
two qrel sets. A passage that survives answer-support filtering keeps its
original relevance grade; a filtered passage receives zero gain. We use
nDCG@10 with linear gain and compare the resulting system orderings with
Kendall's $\tau_b$.

\smallskip
\noindent
\textbf{Answer support changes every leaderboard.}
Relevance and answer support disagree substantially. Among the three BEIR
collections, 63.4\% of SciFact positives satisfy the answer-support
criterion, compared with 43.4\% for TREC-COVID and 18.6\% for NFCorpus.
That difference changes the system ordering on every benchmark.
Table~\ref{tab:rq1} summarizes the result. Kendall's $\tau_b$ between the
original and answer-support-aware rankings ranges from 0.378 to 0.746,
with 6--14 pairwise reversals. The top-ranked system changes on four of the
five collections.

\begin{table}[t]
\centering
\caption{RQ1 results. $\tau_b$ compares the original and
answer-support-aware system orderings. $p_{\mathrm{rand}}$ compares the
observed reordering with matched-random judgment removal. Exposure spread
is the largest between-system difference in the share of original DCG@10
contributed by positives removed by answer-support filtering.}
\label{tab:rq1}
\begin{tabular}{lrrrr}
\toprule
Collection & $\tau_b$ & Rev. & $p_{\mathrm{rand}}$ & Exp. spread \\
\midrule
TREC-COVID & 0.378 & 14 & 0.0243 & 33.9 pp \\
SciFact    & 0.556 & 10 & 0.3120 &  5.5 pp \\
DL2019     & 0.636 & 10 & 0.4783 &  5.4 pp \\
NFCorpus   & 0.733 &  6 & 0.1097 &  7.1 pp \\
DL2020     & 0.746 &  7 & 0.2108 &  5.3 pp \\
\bottomrule
\end{tabular}
\end{table}

\begin{figure}[t]
    \centering
    \includegraphics[width=\columnwidth]{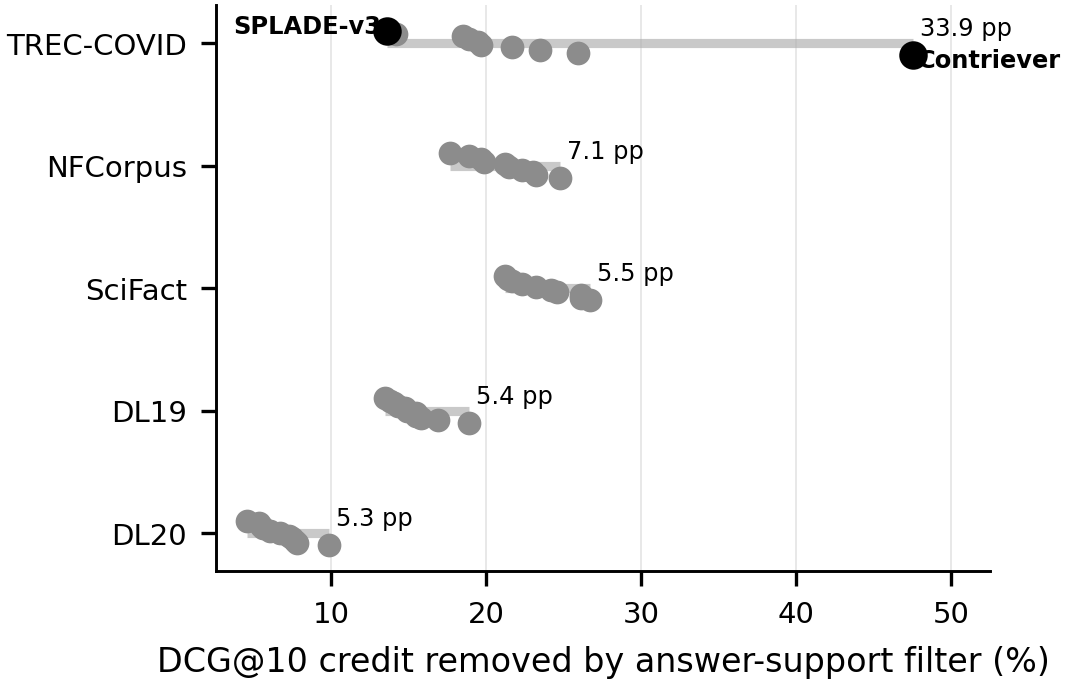}
    \caption{System-selective exposure to relevance--answer-support mismatch.
    Each point represents a retrieval system and shows the share of its
    original evaluated DCG@10 contributed by positive judgments that fail the
    answer-support criterion. TREC-COVID exhibits a 33.9 percentage-point
    spread between systems, compared with at most 7.1 points on the other
    collections.
    }
    \label{fig:system_selective_mismatch}
\end{figure}

\smallskip
\noindent
\textbf{System-selective mismatch.}
A large change in the qrels can itself move a leaderboard. We therefore
compare answer-support filtering with matched-random removal. For each query
and relevance grade, we remove exactly the same number of positive judgments
as answer-support filtering, but choose them randomly. We repeat this
$100{,}000$ times.
TREC-COVID is the only collection whose observed reordering exceeds the
matched-random distribution ($p_{\mathrm{rand}}=0.0243$). The other four
collections have $p_{\mathrm{rand}}=0.1097$--$0.4783$. The amount of
judgment removal does not account for this difference: NFCorpus loses
81.4\% of its positives, compared with 56.6\% for TREC-COVID, yet its
reordering remains consistent with matched-random removal.
The distinguishing factor is how unevenly the removed judgments benefit
competing systems. To measure how strongly each system depends on the judgments being removed,
we define its exposure as the fraction of its original DCG@10 gain contributed
by positives that fail the answer-support criterion. On TREC-COVID, this share is 47.5\% for Contriever and 13.6\%
for SPLADE-v3, a 33.9 percentage-point (pp) difference. The largest spread on
any other collection is only 7.1 points
(Figure~\ref{fig:system_selective_mismatch}).

Across these benchmarks, relevance--evidence mismatch becomes most
consequential when it is \emph{system-selective}. A large amount of mismatch
can leave relative system comparisons largely intact when it affects systems
similarly. When the mismatched judgments benefit competing systems
differently, changing the evidence criterion can substantially change the
leaderboard.

\smallskip
\noindent
\textbf{The effect persists across judges.}
For this judge-sensitivity analysis, GPT-4.1, Qwen3-30B, and Llama-3-8B
evaluate exactly the same query--passage pairs from a shared BM25 top-100
candidate pool. Binary
answer-support agreement yields Gwet's AC1 values of 0.626--0.731 across
collections. More importantly, all 15 judge--collection combinations reorder
the relevance-based leaderboard. All three judges select the same
answer-support-aware winner on three collections, and two of three agree on
the remaining two.
The RQ1 effect therefore persists across all three answer-support judges:
changing the evidence criterion changes retrieval conclusions.

\smallskip
\noindent
\textbf{Takeaway (RQ1).}
Retrieval conclusions are not independent of what the benchmark counts as
useful. Moving from topical relevance to answer-supporting evidence can change
which systems appear better, especially when systems differ in how much their
retrieved relevance actually consists of answer-bearing evidence. The
important issue is therefore not only how much relevance and evidence
disagree, but whether that disagreement is distributed unevenly across
systems. RQ2 asks whether acting on these changed conclusions leads to better
RAG systems.

%% file: 05_optimization_validity.tex
\section{Evidence-Aware System Decisions}
\label{sec:system_decisions}

\S\ref{sec:retrieval_conclusions} showed that answer-support-aware
evaluation can change which retrievers appear better. RQ2 asks whether acting
on that signal leads to better systems. We test two consequential uses:
shaping retriever training and selecting a retriever for downstream RAG.

\smallskip
\noindent
\textbf{Using answer support for retrieval training.}
We first ask whether answer-support judgments provide useful training
supervision.

\textbf{\textit{Training setup.}}
We fine-tune
\nolinkurl{ms-marco-MiniLM-L-6-v2}
with a RankNet objective for three epochs (learning rate
$2\times10^{-5}$), using five query-disjoint folds and dev nDCG@10 for
checkpoint selection. Before constructing the folds, we restrict each
collection to queries for which the BM25 candidate set contains at least one
training positive: a passage with original relevance grade $\geq2$ for
NFCorpus and TREC-COVID, or grade $\geq1$ for binary-qrel SciFact, and
answer-support grade $\geq2$. This yields 87 of 323 NFCorpus queries, all
50 TREC-COVID queries, and 188 of 300 SciFact queries. 
Every retained query
is held out exactly once across the five folds.

Within each query, retrieved assumed negatives are separated into
answer-support grade 0 (C0: no answer-supporting evidence) and grade 1
(C1: partial or insufficient evidence). We compare equal C0/C1 weighting
with a variant that gives C1 negatives $1.5\times$ weight. The matched
control uses the same number of negatives from the same candidate pool,
matched by BM25 rank, but ignores the answer-support distinction.

\textbf{\textit{Training results.}}
Answer-support-aware training does not reliably improve nDCG@10 over the
matched control. With equal weighting, the differences are $+0.0027$ on
NFCorpus (95\% CI: $-0.0050$ to $+0.0108$), $+0.0021$ on TREC-COVID
(95\% CI: $-0.0059$ to $+0.0098$), and $+0.0072$ on SciFact
(95\% CI: $-0.0026$ to $+0.0185$). Weighting C1 more strongly likewise
yields no reliable gain: $+0.0030$, $-0.0035$, and $+0.0069$,
respectively, with every confidence interval including zero. Thus, although
answer support exposes distinctions that conventional relevance collapses,
using those distinctions as hard-negative supervision does not reliably
produce a better retriever.

\smallskip
\noindent
\textbf{From retrieval rankings to system selection.}
We next ask: if answer-support-aware evaluation favors a different retriever,
does choosing that retriever improve downstream answers?

\textbf{\textit{Experimental setup.}} We use the official TREC RAG 2025 retrieval task. Of its 22 judged topics,
four are reserved for prompt calibration and excluded from all reported
results, leaving 18 evaluation topics. We begin from the 46 official retrieval submissions. Forty satisfy our
prespecified eligibility requirements: at least 95\% coverage of the
evaluation topics, retrieval depth of at least 100, valid MS MARCO v2.1
segment identifiers with no within-topic duplicates, a publicly documented
automatic retrieval method with no manual alteration of test-topic rankings,
top-100 segment identifiers that all resolve to corpus text, and successful
evaluation by the released TREC retrieval scorer without modification.

We then construct a fixed downstream-analysis roster with one representative
run per team, avoiding over-representation of teams with multiple submissions. We use a team-declared primary run when available;
otherwise, we choose that team's eligible run with the highest official
original-qrel retrieval score. This de-duplication yields 11 systems and is
distinct from the cross-fitted retriever-selection experiment below. Neither
answer-support judgments nor downstream answer quality is used to construct
the roster.


For each system, we evaluate retrieval with nDCG@30 under either the original
TREC RAG 2025 relevance judgments or answer-support-aware judgments. For RAG25,
the same claim-based instrument as in
\S\ref{sec:retrieval_conclusions} uses the full topic narrative as the
information need, with GPT-4.1 as the primary judge. An originally relevant
passage contributes gain only when its answer-support grade is at least 2.

\textbf{\textit{Generation conditions.}}
We generate an answer for each retriever--topic pair using Gemma3-27B and
the passages returned by that retriever. We evaluate two fixed generation
strategies. The \emph{Standard Grounded} prompt asks the model to answer the
query as fully as the supplied passages allow, use only information
supported by those passages, cite every factual statement, and explicitly
identify requested information that the passages do not support. The
\emph{Coverage-Disciplined} prompt retains these requirements but additionally
encourages broader coverage of distinct supported facts, nonredundant
citations, atomic claims, explicit treatment of supported negative findings,
and representation of disagreement among passages.

We use four held-out topics to choose the primary generation condition before
analyzing the remaining 18 topics. Under a prespecified rule, Coverage-Disciplined would replace Standard Grounded only if it improved
coverage of fully supported vital nuggets by at least $0.02$ while reducing
citation precision by no more than $0.01$. Although it improved citation precision, it reduced vital-nugget coverage by
$0.027$; Standard Grounded therefore remains the primary condition, with
Coverage-Disciplined retained as a prespecified sensitivity condition.

\textbf{\textit{Answer-quality evaluation.}} We evaluate the generated answers with RAGDoll's Nuggetizer, which compares
an answer against the information nuggets defined for each TREC RAG topic.
Our primary answer-quality measure is
\texttt{strict\_vital\_score}: it considers only nuggets designated as vital
and gives credit only when a vital nugget is fully supported by the generated
answer; partial support receives no credit. Qwen3-30B provides the primary
Nuggetizer judgments. Claude Sonnet~5 independently evaluates the same
generated answers as an evaluator sensitivity.

\textbf{\textit{Cross-fitted system selection.}}
Within the fixed 11-system roster, retriever selection is cross-fitted over
the 18 evaluation topics. In each fold, each retrieval criterion selects the
system with the highest mean nDCG@30 on the training topics, and we measure
that system's answer quality on the held-out topics. Thus, held-out topics do
not participate in the fold-specific retriever selection, and downstream
answer quality never participates in selection.

\textbf{\textit{Selection results.}} At the system-ranking level, answer-support-aware evaluation appears
promising. With Qwen evaluating answer quality, Kendall's $\tau_b$ between
retrieval effectiveness and downstream answer quality increases from
$0.2364$ under conventional relevance to $0.3455$ under answer-support-aware
evaluation ($\Delta\tau=+0.1091$). With Claude evaluating the same generated
answers, it increases from $0.2364$ to $0.4182$
($\Delta\tau=+0.1818$). Both changes are directionally favorable, although
their topic-bootstrap confidence intervals include zero.

The held-out decision tells a different story. Under the primary
Standard Grounded generation condition, selecting a retriever by
answer-support-aware rather than conventional relevance evaluation does not
improve downstream answer quality. With Qwen, the mean held-out difference is
$-0.0065$ (95\% CI $[-0.0617,+0.0497]$, $p=0.828$); with Claude it is
$+0.0018$ (95\% CI $[-0.0235,+0.0260]$, $p=0.902$). The null result is not
because the two evaluation criteria make the same decision: in the
prespecified five-fold assignment, they select the same retriever in only one
fold. Answer-support-aware evaluation therefore changes which system is
chosen, but under the primary generation condition that changed choice does
not improve held-out answers.

\smallskip
\noindent
\textbf{Decision value depends on evidence use.}
We now repeat the same selection experiment under the
Coverage-Disciplined generation condition. The retrieval systems,
answer-support judgments, retrieval metric, cross-fitted selection procedure,
and Gemma3-27B generator remain unchanged; only the instructions governing how
the generator uses the retrieved evidence differ.

Answer-support-aware system selection improves held-out
answer quality. The gain is $+0.0583$ with Qwen
(95\% CI $[+0.0202,+0.1069]$, $p=0.007$) and $+0.0324$ with Claude
(95\% CI $[+0.0072,+0.0588]$, $p=0.030$). Selection regret---the gap between the selected retriever's answer quality
and that of the best available retriever on the held-out topics---decreases
by $0.0569$ and $0.0311$, respectively.

\begin{table}[t]
\centering
\caption{Held-out change in downstream answer quality when retrievers are
selected using answer-support-aware rather than original relevance
evaluation. Positive values favor answer-support-aware selection.}
\label{tab:rq2_selection}
\scalebox{0.8}{
\begin{tabular}{llrr}
\toprule
Generation & Evaluator & $\Delta$ answer quality & 95\% CI \\
\midrule
Standard Grounded
    & Qwen3-30B & $-0.0065$ & $[-0.0617,+0.0497]$ \\
Standard Grounded
    & Claude Sonnet 5 & $+0.0018$ & $[-0.0235,+0.0260]$ \\
Coverage-Disciplined
    & Qwen3-30B & $\mathbf{+0.0583}$ & $[+0.0202,+0.1069]$ \\
Coverage-Disciplined
    & Claude Sonnet 5 & $\mathbf{+0.0324}$ & $[+0.0072,+0.0588]$ \\
\bottomrule
\end{tabular}
}
\end{table}

The contrast identifies the central RQ2 result. Answer-support-aware
evaluation can change retrieval rankings and move them closer to downstream
answer-quality rankings without providing a universally better
system-selection rule. Under Standard Grounded, the changed decision provides
no held-out benefit. Under Coverage-Disciplined, the same retrieval criterion
and selection procedure produce a reliable gain. The selection value of
answer-support-aware evaluation, therefore, depends on how the downstream
generator is instructed to use the evidence that the criterion favors.

\smallskip
\noindent
\textbf{Takeaway (RQ2).}
A better evidence signal is not automatically a better decision rule.
Answer-support distinctions do not reliably improve retrieval training in our
experiments, while answer-support-aware system selection improves held-out
answers only under one generation regime. The value of an evaluation signal
is therefore partly a property of the pipeline and decision in which it is
used, not of the metric in isolation. RQ3 asks whether retrieval scores
themselves nevertheless generalize as predictors of downstream answer quality.

%% file: 06_downstream_validity.tex
\section{Predictive Validity of Evidence-Aware Scores}
\label{sec:predictive_validity}

RQ2 showed that answer-support-aware evaluation can align more closely with
downstream system rankings without consistently supporting better system
selection. RQ3 asks whether retrieval scores themselves nonetheless provide
useful information about answer quality on unseen topics.

\textbf{\textit{Predictive setup.}}
We use the same 18 topics and 11 systems as in
\S\ref{sec:system_decisions}. For each topic--system pair, we observe
original nDCG@30 ($O$), answer-support-aware nDCG@30 ($G$), and downstream
answer quality. We fit three ordinary least-squares models: $O$ predicts
answer quality from conventional retrieval effectiveness, $G$ uses
answer-support-aware effectiveness, and $OG$ uses both. Thus, $G$ tests
whether answer-support-aware evaluation is a better predictor than
conventional relevance, while $OG$ tests whether it adds predictive
information beyond conventional relevance.

We evaluate transfer with leave-one-topic-out cross-validation. Within each
training fold, retrieval features and answer quality are centered within
topic before estimating slopes. For the held-out topic, only retrieval
features are centered using that topic's retrieval scores; no held-out answer
score is used. We report pooled cross-validated $R^2$ relative to the pooled
held-out mean, so negative values indicate greater squared prediction error
than that baseline. Predictive validity therefore requires a relationship
learned across training topics to generalize to systems on an unseen topic.

\textbf{\textit{Results.}}
Under the primary Standard Grounded condition, none of the retrieval features
provides useful out-of-topic prediction. With Qwen evaluation,
$R^2_{\mathrm{CV}}$ is $-0.0726$ for $O$, $-0.0800$ for $G$, and
$-0.0802$ for $OG$. With Claude, the corresponding values are $-0.0495$,
$-0.0491$, and $-0.0503$. Thus, answer-support-aware retrieval scores are no
more predictive than conventional relevance in the primary condition, and
combining the two signals does not help.

The linear result is consistent across the full analysis: across two
generation prompts, five generator--grounding-judge configurations, and two
answer evaluators, all 60 leave-one-topic-out OLS models have negative
cross-validated $R^2$. Adding answer-support-aware retrieval to 
relevance improves $R^2_{\mathrm{CV}}$ numerically in only four of the 20
conditions, and all four models remain negative.

We also test whether this failure is an artifact of assuming a linear
relationship. Repeating the same leave-one-topic-out protocol with
low-capacity cubic-spline regression yields negative $R^2$ in 55/60 tested
cells, while monotone isotonic regression is negative in 56/60. The few
positive nonlinear values are negligible ($R^2\leq0.007$) and occur only
under Claude evaluation with Coverage-Disciplined Gemma3 generation. Thus,
allowing smooth or monotone nonlinear relationships does not reveal a robust
transferable relationship between retrieval effectiveness and answer quality.

\begin{figure*}[t]
    \centering

    \begin{minipage}[t]{0.30\textwidth}
        \vspace{0pt}
        \centering
        \includegraphics[width=\linewidth]{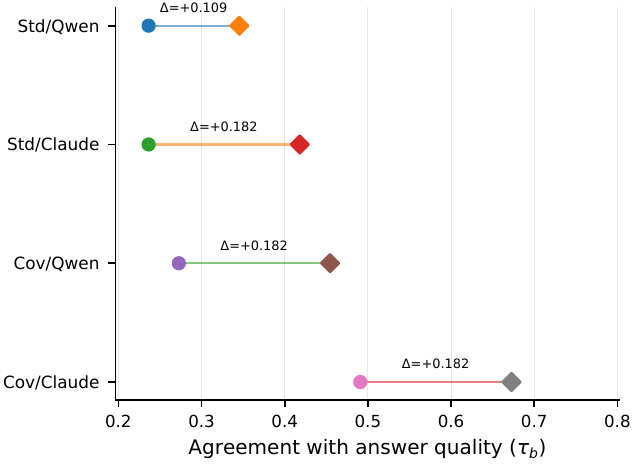}
        \vspace{-1mm}
        {\small\textbf{(a)} Ranking alignment
        ($\circ$ O, $\diamond$ G)}
    \end{minipage}
    \hfill
    \begin{minipage}[t]{0.31\textwidth}
        \vspace{0pt}
        \centering
        \includegraphics[width=\linewidth]{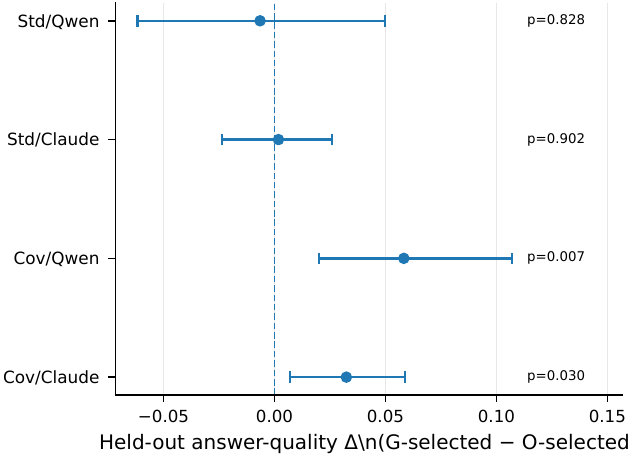}
        \vspace{-1mm}
        {\small\textbf{(b)} Held-out selection value}
    \end{minipage}
    \hfill
    \begin{minipage}[t]{0.35\textwidth}
        \vspace{0pt}
        \centering
        \includegraphics[width=\linewidth]{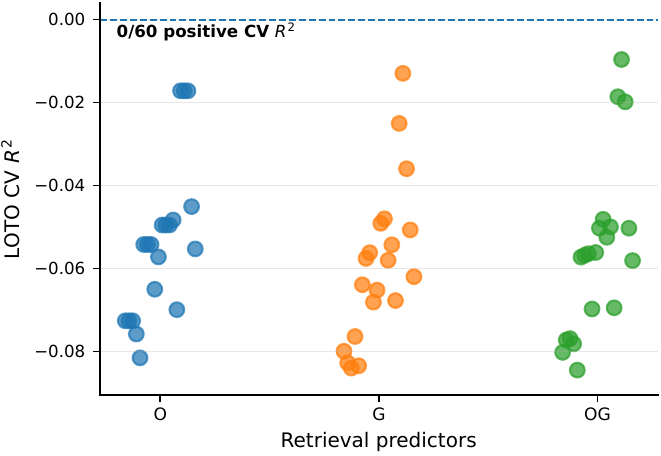}
        \vspace{-1mm}
        {\small\textbf{(c)} Out-of-topic prediction}
    \end{minipage}
    \caption{\textbf{Downstream value of answer-support-aware retrieval.}
(a) Answer-support-aware scores show higher observed alignment with answer
quality. (b) This does not consistently improve held-out system selection:
gains appear under Coverage-Disciplined but not Standard Grounded generation.
(c) Neither original nor answer-support-aware scores provide transferable
out-of-topic prediction; all 60 OLS models have negative cross-validated
$R^2$, with nonlinear sensitivities yielding the same overall conclusion.}

    \label{fig:downstream_validity}
\end{figure*}

Taken together with RQ2, these results separate three distinct uses of the
same retrieval signal: ranking alignment, held-out system selection, and
out-of-topic prediction (Figure~\ref{fig:downstream_validity}).
Answer-support-aware evaluation can move the aggregate system ordering closer
to the downstream ordering without becoming a reliable selection rule under
every generation regime or a score that robustly predicts answer quality on
unseen topics.

\smallskip
\noindent
\textbf{Takeaway (RQ3).}
RQ1--RQ3 reveal a validity-composition problem. Answer support can be more
directly aligned with answer-bearing evidence without every downstream use of
that signal becoming valid in turn. It changes retrieval conclusions; its
value for retrieval training is not established; its value for system
selection depends on the generation regime; and its scores provide no robust
transferable prediction of answer quality on unseen topics. Better alignment
at one stage therefore does not establish the validity of the downstream use.
This motivates a more direct test. RQ4 leaves retrieval scores behind and
intervenes on the evidence supplied to the generator itself. We test whether
answer-support judgments can identify evidence worth preserving and, if so,
whether changing that evidence changes the quality of the generated answer.

%% file: 07_intervention_validity.tex
\section{Does Better Evidence Improve RAG?}
\label{sec:intervention_validity}

RQ3 found no robust out-of-topic prediction from retrieval scores. RQ4 therefore intervenes directly on the retrieved context, asking whether a human-validated shift toward answer-bearing evidence yields an evaluator-robust improvement in answer quality.

\textbf{\textit{Controlled evidence intervention.}}
RQ1--RQ3 asked what happens when answer support is used to evaluate,
optimize, select, or predict retrieval. RQ4 asks a more direct question:
\emph{does the signal actually identify evidence that is more useful to the
generator?} We therefore hold the retriever fixed and use answer support only
to change which of its retrieved passages reach the generator.

For each held-out topic, we begin with the retriever that would have been
selected using the original RAG25 retrieval judgments. Within each of
five folds, we select the retriever with the highest nDCG@30 on
that fold's training topics and apply it to the held-out topics. This simulates the context that conventional benchmark evaluation would have
led us to use while keeping the held-out topics out of the fold-specific
retriever selection decision.

We then construct an answer-support-filtered version of the retrieved context
using the judgments defined in \S\ref{sec:framework}. We remove passages that
fall below the answer-support criterion and backfill, in ranked order, with
lower-ranked passages that satisfy it until reaching 20 passages or the fixed
12,000-token budget. This directly tests whether preferentially supplying
answer-supporting passages gives the generator better evidence.

Simply changing which passages reach the generator can itself affect
generation, so we compare this intervention against an equal-magnitude
matched-random control. For each passage displaced by answer-support
filtering, the control removes a passage matched on original TREC relevance
grade category, retrieval-rank band, and passage-length quartile. When an
exact match is unavailable, matching follows a fixed relaxation sequence:
drop the length quartile, merge adjacent rank bands, retain only the relevance
category, and finally use the global candidate pool. The context is then
backfilled from the same ranked list in original order. We generate 20
matched-random controls per topic using seeds 20260731--20260750 and average
their answer outcomes within topic before comparison. Thus, both conditions
perturb contexts from the same retriever under the same context budget, while
answer support determines which passages the intervention preferentially
preserves.

\textbf{\textit{Human validation of the mechanism.}}
Because the intervention is defined by GPT-4.1 answer-support judgments, we
validate two distinct links before interpreting its downstream effect. \textbf{H1} asks whether the answer-support distinction itself is
human-recognizable; \textbf{H2} asks whether using that distinction to filter
context actually moves the supplied evidence in the intended direction.

We construct two independently sampled 60-passage studies from the retrieved
contexts of the same 18 TREC RAG 2025 evaluation topics used in the preceding
experiments, with 3--4 passages from every topic. H1 is deliberately a
challenge sample rather than a prevalence estimate: 30 passages come from
cases where GPT-4.1, Qwen, and Llama agree on the answer-support boundary and
30 from cases where the judges disagree, with both GPT-4.1-positive and
GPT-4.1-negative decisions represented. H2 instead samples passages according
to the decisions made by answer-support filtering and the prespecified
matched-random control: passages uniquely removed by either policy, together
with passages retained by both. The two studies overlap on 13 passages,
yielding 107 unique passages for annotation.

Two independent annotators inspect the original passage text and label each
passage as containing no meaningful evidence (0), related but insufficient
information (1), or answer-supporting evidence (2). They do not see the
decomposed claims, study stratum, automated judgments, or filtering decision. Across all passages, exact agreement is
93.5\%, with Cohen's $\kappa=0.660$ and Gwet's AC1 $=0.928$; at the binary
0/1-versus-2 evidence boundary, AC1 is $0.919$.

For \textbf{H1}, GPT-4.1 agrees with the two annotators on 73.3\% and
68.3\% of the 60 challenge cases. The disagreement is strongly asymmetric.
Both annotators classify all 16 passages labeled answer-supporting by all
three automated judges as evidence-bearing. By contrast, among the 14
passages labeled non-supporting by all three, 71.4\% and 78.6\% are still
judged evidence-bearing by humans. The automated boundary therefore captures
a human-recognizable evidence distinction, but GPT-4.1's negative decisions
are comparatively conservative and should not be treated as ground
truth.

For \textbf{H2}, we test the intervention itself using a prespecified
matched-random control (seed 20260731). Passages preserved by
answer-support filtering but removed by this control are judged
evidence-bearing in 14/15 cases by both annotators. Passages removed by
answer-support filtering but retained by the random control are
evidence-bearing in 11/15 and 12/15 cases, differences of 20.0 and 13.3
percentage points. Thus, although the filter does not perfectly separate
useful from useless passages, it preferentially preserves evidence that
humans recognize as answer-bearing.

\textbf{\textit{Downstream effect.}}
Having established that answer-support filtering preferentially preserves
human-recognized evidence, we next ask whether that improvement carries
through to the generated answer. Under the primary Qwen3-30B evaluator, and
using the same primary downstream measure as in RQ2--RQ3,
\texttt{strict\_vital\_score} (the fraction of vital nuggets fully supported
by the answer), answer-support filtering increases the score by $+0.0289$
relative to the matched-random control (95\% CI
$[+0.0051,+0.0518]$), with improvements on 13 of 18 topics
($p=0.0305$, exact sign-flip test). Neither \texttt{strict\_all\_score}
(full support across all nuggets) nor \texttt{hard\_recall} (strict
citation-support recall) shows the same reliable improvement. The positive
effect is therefore specific to the primary measure rather than a uniform
gain across the evaluated answer dimensions.

\begin{figure*}[t]
    \centering
    \begin{minipage}[t]{0.48\textwidth}
        \vspace{0pt}
        \centering
        \includegraphics[height=0.245\textheight]{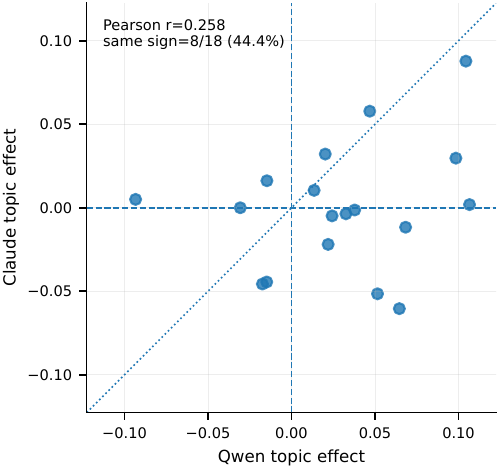}
        \vspace{1mm}
        \parbox[t]{0.94\linewidth}{
            \centering
            \small\textbf{(a)} Topic-level evaluator agreement
        }
    \end{minipage}
    \hfill
    \begin{minipage}[t]{0.48\textwidth}
        \vspace{0pt}
        \centering
        \includegraphics[height=0.245\textheight]{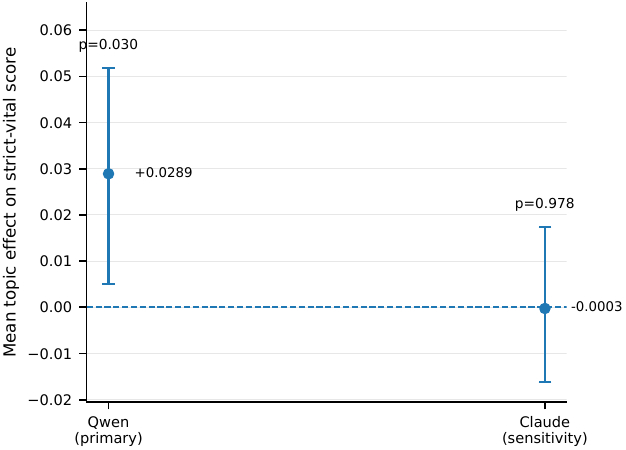}
        \vspace{1mm}
        \parbox[t]{0.94\linewidth}{
            \centering
            \small\textbf{(b)} Aggregate treatment effect
        }
    \end{minipage}
    \caption{\textbf{Evaluator sensitivity of the RQ4 treatment effect.}
(a) Qwen and Claude show weak topic-level agreement
($r=0.258$; same sign on 8/18 topics).
(b) The aggregate conclusion changes: Qwen finds a positive intervention
effect, while Claude estimates essentially none.}
    \label{fig:evaluator_sensitivity}
\end{figure*}

\textbf{\textit{Evaluator sensitivity.}}
The remaining question is whether this downstream conclusion is robust to how
answer quality itself is measured. We therefore evaluate the exact same
generated answers with Claude Sonnet~5, designated as an independent evaluator
sensitivity before the treatment-effect analysis. Claude does not reproduce
the Qwen result: the mean effect is $-0.0003$ (95\% CI
$[-0.0162,+0.0173]$), with 8 wins, 1 tie, and 9 losses across the 18 topics
($p=0.9780$). Thus, at the aggregate level, the same intervention supports a
positive treatment effect under Qwen and essentially no effect under Claude
(Fig.~\ref{fig:evaluator_sensitivity}(b)).

The disagreement is not simply a consequence of one $p$-value crossing a
significance threshold. Across the 18 topics, the treatment effects estimated
by the two evaluators correlate only $r=0.258$ and agree in sign on 8 topics
(Fig.~\ref{fig:evaluator_sensitivity}(a)). The evaluator dependence is
therefore visible not only in the aggregate estimate, but in how the two
evaluators assess the intervention topic by topic. Even after human validation
confirms that answer-support filtering moves the context toward more
answer-bearing evidence, whether that change appears to improve the final
answer depends substantially on how answer quality is evaluated.

\smallskip
\noindent
\textbf{Takeaway (RQ4).}
RQ4 isolates the final break in the validity chain. Human judgments confirm
that the intervention moves the retrieved context toward answer-bearing
evidence, so the downstream disagreement cannot be explained simply by the
intervention failing to manipulate what it intended to manipulate. Yet the
same contexts and the same generated answers support an improvement under
one evaluator and essentially no effect under another. The answer evaluator
is therefore part of the validity of the downstream conclusion, not merely a
reporting device.
Across RQ1--RQ4, this exposes a validity-composition problem: evidence that a
measurement or intervention is meaningful at one stage does not automatically
validate the comparison, decision, prediction, or conclusion drawn at the
next.

%% file: 08_discussion.tex
\section{Discussion}
\label{sec:discussion}

Our experiments show that ``better RAG evaluation'' is not a single property.
An evaluation signal may better reflect the evidence needed for generation,
change which systems appear better, support some downstream decisions, and
fail at others. Answer support exhibits exactly this pattern. It changes
retrieval conclusions and identifies evidence that humans recognize as
answer-bearing, yet it does not reliably improve retrieval training, its value
for system selection depends on the generation regime, and its retrieval
scores do not robustly predict answer quality on unseen topics. Even after a
direct intervention improves the supplied evidence, the conclusion that the
resulting answers improved depends on the answer evaluator. These results
separate several properties that are often treated as if they followed from
one another: construct alignment, system-comparison validity, decision
validity, predictive validity, and robustness of outcome measurement.

\textbf{\textit{Evaluation utility is relational.}}
Two patterns help explain why a more meaningful retrieval signal does not have
uniform downstream value. First, the amount of disagreement between relevance
and answer support alone is not sufficient to determine how consequential the
change will be. NFCorpus loses a large share of its original positive
judgments, yet that loss is distributed relatively evenly across systems. In
TREC-COVID, by contrast, competing retrievers differ much more in their
exposure to judgments that fail the answer-support criterion, and changing the
criterion has a correspondingly larger effect on their relative ordering.
This suggests that the consequence of an evaluation change depends not only
on what the criterion measures, but also on how its disagreements are
distributed across the systems being compared.

Second, the value of the resulting comparison depends on what happens after
retrieval. The same answer-support-aware selection procedure provides no
reliable held-out benefit under Standard Grounded generation but improves
answer quality under Coverage-Disciplined generation. The retrieval criterion,
systems, and selection procedure are unchanged; what changes is how the
generator is instructed to use the retrieved evidence. Evaluation utility is
therefore not solely a property of a metric. It arises from the interaction
between the evaluation criterion, the systems being compared, the downstream
component consuming their outputs, and the decision the evaluation is meant
to support.

\textbf{\textit{Validation should follow intended use.}}
This distinction matters because retrieval evaluation increasingly functions
as decision infrastructure. Offline metrics are not used only to describe
systems; they guide optimization, model selection, and deployment. Validation
should therefore test the use for which an evaluation signal is intended.
If a metric is used to select a retriever, its value should be tested through
held-out system selection rather than inferred from stronger correlation with
downstream quality. If retrieval effectiveness is used as a proxy for
end-to-end performance, the relationship should transfer to unseen topics
rather than hold only within the observed benchmark. And when an automated
answer evaluator is used to determine whether an intervention improved a
system, robustness should be assessed at the level of that scientific
conclusion, not only through item-level agreement between evaluators.
The broader methodological implication is that meta-evaluation should mirror
the downstream inference or decision that the metric will actually support.

\textbf{\textit{Beyond passage-level answer support.}}
Our experiments also expose a limitation of passage-level evidence judgments.
A generator consumes an evidence set, not isolated passages. Individually
answer-supporting passages may be redundant, collectively omit important
aspects of the information need, or consume limited context budget with
overlapping evidence. Conversely, a passage with modest value in isolation
may be useful because it complements what has already been retrieved.
Evidence-set utility may therefore depend on coverage, redundancy,
complementarity, contradiction, and context budget in addition to the
answer-support value of individual passages. The stronger system-selection
result under Coverage-Disciplined generation is consistent with the idea that
the downstream value of retrieved evidence depends on how effectively the
generator exploits such structure, although our experiments do not establish
that mechanism.

Taken together, these results suggest a broader direction for RAG evaluation.
The field is right to move beyond topical relevance toward evidence that is
more closely connected to downstream needs, but richer constructs alone do
not resolve the evaluation problem. The next question is not only whether a
metric captures something meaningful, but whether it supports the particular
comparison, decision, prediction, or experimental conclusion for which it is
used. In multi-stage RAG systems, those transitions are themselves objects of
validation.

%% file: 09_conclusion.tex
\section{Conclusion}
\label{sec:conclusion}

We asked how far the benefits of answer-support-aware retrieval evaluation
carry through the RAG pipeline. Across retrieval comparison, training, system
selection, prediction, and a direct evidence intervention, the answer is: not
automatically. Answer support changes retrieval conclusions and identifies
evidence that humans recognize as more answer-bearing, yet its downstream
value depends on the decision being made, how the generator uses the retrieved
evidence, and how the resulting answers are evaluated.

We call this the \emph{validity-composition problem}: evidence that an
evaluation is meaningful for one use does not automatically validate the next
inference or decision built from it. As RAG evaluation increasingly guides
system development and deployment, validity must therefore be established for
the particular use an evaluation signal is intended to support rather than
assumed to propagate through the pipeline.

%% file: 10_ethics.tex
\section*{Ethical Considerations}

This work uses public information-retrieval benchmarks and does not involve
private user data or deployment on real users. Human annotation is limited to
judging the evidential value of retrieved passages. Because several analyses
rely on LLM-based judgments, these judgments may inherit model-specific biases
or systematic errors; we therefore evaluate cross-judge robustness and include
independent human validation for the central evidence intervention. More
broadly, our results caution against treating automated RAG metrics as
decision-neutral: when such metrics guide system selection, optimization, or
deployment, their errors can propagate into consequential engineering
decisions. We therefore view transparency about evaluator choice, validation
against the intended use, and reproducible reporting of evaluation procedures
as important safeguards for responsible RAG evaluation.